\documentclass[12pt]{iopart}
\usepackage{iopams}
\usepackage{graphicx}

\newcommand{\eps}{\varepsilon}
\newcommand{\la}{\langle}
\newcommand{\ra}{\rangle}
\newcommand{\lla}{\left\langle}
\newcommand{\rra}{\right\rangle}
\newcommand{\da}{^{\dagger}}
\newcommand{\diff}{{\rm d}}
\newcommand{\vp}{\boldsymbol{p}}
\newcommand{\vk}{\boldsymbol{k}}
\newcommand{\vu}{\boldsymbol{u}}
\newcommand{\vr}{\boldsymbol{r}}

\newcommand{\N}{\mathbb {N}}
\newcommand{\R}{\mathbb {R}}
\newcommand{\cL}{ {\cal L} }
\newcommand{\cP}{ {\cal P} }
\newcommand{\cQ}{ {\cal Q} }
\newcommand{\Order}[1]{ {\cal O} \left( #1 \right) }
\newcommand{\Op}[1]{{\sf #1}} 
\newcommand{\oH}{ \Op{H} } 
\newcommand{\oA}{ \Op{A} } 
\newcommand{\oE}{ \Op{E} } 
\newcommand{\oB}{ \Op{B} } 
\newcommand{\oa}{ \Op{a} } 
\newcommand{\ok}{ \Op{k} } 
\newcommand{\og}{ \Op{g} } 
\newcommand{\op}{ \Op{p} } 
\newcommand{\ox}{ \Op{x} } 
\newcommand{\ovp}{ \boldsymbol{\sf p}} 
\newcommand{\ovr}{ \boldsymbol{\sf r}} 

\newcommand{\re}{\Re\mathfrak{e}\, }
\newcommand{\im}{\Im\mathfrak{m}\,}

\begin{document}

\title{Master Equation for the Motion of a Polarizable Particle in a Multimode Cavity}
\author{Stefan Nimmrichter$^1$, Klemens Hammerer$^2$, Peter Asenbaum$^1$, Helmut Ritsch$^3$, and Markus Arndt$^1$}
\address{$^1$ Faculty of Physics, University of Vienna, Boltzmanngasse 5, 1090 Vienna, Austria}
\ead{stefan.nimmrichter@univie.ac.at}
\address{$^2$ Institute for Quantum Optics and Quantum Information of the Austrian Academy of Sciences, 6020 Innsbruck, Austria}
\address{$^3$ Institute for Theoretical Physics, Technikerstra{\ss}e 25, 6020 Innsbruck, Austria}

\date{\today}

\begin{abstract}
We derive a master equation for the motion of a polarizable particle weakly interacting with one or several strongly pumped cavity modes.
We focus here on massive particles with a complex internal structure such as large molecules and clusters, for which we assume a linear scalar polarizability mediating the particle-light interaction. The predicted friction and diffusion coefficients are in good agreement with former semiclassical calculations for atoms and small molecules in weakly pumped cavities, while the current rigorous quantum treatment and numerical assessment sheds light on the feasibility of experiments that aim at optically manipulating beams of massive molecules with multimode cavities.
\end{abstract}

\submitto{\NJP}

\maketitle

\section{Non-resonant, cavity-assisted laser cooling}
Laser cooling of atoms has been an important driving force in quantum optics throughout the last three decades~\cite{Varenna1993a,Wieman1999a,Weidemuller2003a}. It has led to great advances in atomic spectroscopy and atomic clocks~\cite{Adam2003a}, atom interferometry~\cite{Cronin2009a}, and quantum sensors~\cite{Gustavson1997a,Peters1999a}, and it even paved the way towards the new physics of quantum degenerate gases~\cite{Anderson1995a,Davis1995b}. The optical manipulation of atoms has become a routine tool in many experiments world-wide. A limitation of laser cooling arises, however, when we need to address systems with a more complex internal structure than atoms. Large molecules, atomic clusters or solids in general do not exhibit suitable dipole-allowed transitions that could be used for running a closed cycle of absorption and spontaneous reemission to the initial state, as required for the controlled dissipation of kinetic energy into a  photonic  emission. Laser cooling of particle motion based on spontaneous light forces is therefore mostly precluded.

It has, however,  been realized that even then optical manipulation and cooling can be achieved~\cite{Horak1997a}  when it is done on particles in a high-finesse optical cavity that preferentially redirects any incident scattered radiation because of the enhanced mode density in a given direction. Blue-shifted Rayleigh scattering of light at a particle that is moving along the cavity axis may thus open a channel for damping the particle's kinetic energy.
The cooling rate then depends primarily on the real part of the particle's off-resonant polarizability $\chi$ as well as on the light intensity and its frequency detuning with respect to the cavity resonance. In a standard setting the incident laser light is red-detuned  ($\omega_L< \omega_C$) with respect to a cavity resonance such that the Doppler shift of a particle moving towards one of the cavity mirrors will support the preferential emission into the cavity mode. The combination of a blue-shifted photon recoil with a velocity-dependent enhancement of the scattering rate leads to a friction force that effectively decelerates the particle.

Resonator-assisted optical cooling has been theoretically proposed for the strong coupling regime~\cite{Horak1997a} and extended towards the collective cooling of a cloud of particles~\cite{Gangl1999a,Vuletic2000a,Black2003a,Asboth2004a,Beige2005a} or the cooling of complex molecules, too~\cite{Lev2008a,Salzburger2009a}. The schemes were extended from the one-dimensional case with different pumping configurations~\cite{Domokos2002a,Kruse2003b,Nagy2006a} to the three-dimensional case, still using a single cavity~\cite{Vuletic2001a}. Even a lowering of the internal temperature of the particles was studied in a cavity-mediated environment~\cite{Morigi2007a}.

As of today, the idea of non-resonant cavity-assisted optical cooling has been experimentally demonstrated for individual atoms~\cite{McKeever2003a,Maunz2004a} as well as for cold atomic ensembles. Collective effects in the interaction between an atomic cloud and the cavity field led to the discovery of collective recoil lasing~\cite{Kruse2003a,Chan2003a}.

A related field has been opened recently in a complementary setting: The laser-assisted cooling of micromechanical oscillators~\cite{Kippenberg2008a,Marquardt2009a} shares much of the physics and the formalism with cavity-assisted cooling of atoms~\cite{Schulze2010a}. The manipulation of cantilevers may thus be understood as a specific form of a more general scheme in which one may either study the free motion of individual atoms or the bending of a micromechanical cantilever that is restricted to a harmonic oscillation close to the quantum mechanical ground state. As tailor-made micromechanical oscillators couple to the light very efficiently and as they withstand high laser intensities, the cooling powers in such experiments were shown to be substantial (see \cite{Aspelmeyer2008}, and references therein).

A number of experiments were already successful at both ends of the mass scale, either with atoms or using small solids. It is, however, still an open challenge to apply cavity cooling schemes to freely moving mesoscopic objects. Such nanoparticles could for instance be the C$_{60}$-fullerenes~\cite{Arndt1999a}, heavier biomolecules~\cite{Hackermuller2003a} or large silica nanospheres~\cite{Chang2010a,Romero-Isart2010a} which have recently been used or proposed for quantum superposition experiments. The implementation of efficient cooling and slowing methods will be a key issue in all those experiments.

On the other hand, most of these examples will operate with objects that have broad optical lines and low oscillator strengths, without being large enough to form a mirror for visible or even infrared light. All these cases have therefore to be regarded in the weak-coupling limit~\cite{Hechenblaikner1998a}, as outlined below.

In this article we present a full quantum description of the motion of such a massive polarizable particle under the influence of a strong laser field and a possibly large manifold of high-finesse cavity modes.
This reproduces known results from former semiclassical derivations but also exceeds those works, as we provide experimental case studies for the motional damping of various nanocluster particles and a proper treatment of the strong pump enhancement of the coupling to the cavity modes.
We furthermore show how the presence of the multimode cavity leads to a velocity damping effect acting on the particle, which increases with the number of accessible high-finesse modes. An explicit calculation of this effect is carried out on the basis of a realistic confocal cavity setup with an accessible manifold of $\sim 10^3$ degenerate modes.
We predict an enhancement of the optical friction force by more than two orders of magnitude when compared to a single-mode cavity.
The experimental parameters are assessed using clusters with masses up to $10^5~$amu, where we also include the absorption and the free-space scattering of photons into our model.

In the first section we start by deriving a master equation for the motion of a polarizable particle in a pumped cavity mode. The explicit derivation is carried out with a single particle in a single cavity mode. We then generalize the model to a one-dimensional treatment of $N$ particles in $M$ modes. In Section \ref{sec:phasespace} the master equation is translated into phase space in order to identify the effective friction force and the diffusion terms that act on the particle to lowest order in the photon recoil. An explicit formula for the velocity dependence of the force beyond linear friction is presented there. We incorporate absorption and Rayleigh scattering into our model in Section \ref{sec:MEabsscatt} and discuss its validity with respect to experimental realizations with large particles, such as atomic clusters. The results are assessed in a parameter regime which is realistic for those mesoscopic particles.

\section{Master equation for a particle in a pumped cavity} \label{sec:me}
We start by considering a polarizable particle placed in a single-mode high-finesse cavity which is characterized by its resonance frequency $\omega_c$ and the photon loss rate $2\kappa$.
A laser at frequency $\omega_p$ pumps the resonator continuously.

The coupling between the particle and the cavity field is mediated
through the off-resonant scalar electric polarizability $\chi$ --
which we assume to be \textit{real, positive} and constant within the frequency range between $\omega_p$ and $\omega_c$ -- and the quadrature components of the quantized electric field operator
\begin{equation}
 \oE^{+} (\vr) = \sqrt{\frac{\hbar \omega_c}{2 \eps_0 V_c}} f (\vr) \oa\da, \quad \oE^{-} (\vr) = \left( \oE^{+} (\vr) \right)\da.
\end{equation}
We denote operators by non-serif letters throughout this article. Here, $V_c$ is the mode volume and $f(\vr)$ the mode function fulfilling the Helmholtz equation $\Delta f(\vr) = - k^2 f(\vr)$ with the wave number $k=2\pi /\lambda$. For a cavity with widely extended plane-parallel mirrors the mode function is simply  $f(x)=\cos k x$. For more practical finite cavities and curved mirrors the transverse beam profile is closer to Gaussian and the mode function is three-dimensional, including both the mode profile and phase shifts related to the wavefront curvature~\cite{Hodgson2005}.

The interaction between the particle and the cavity is described by the field energy of the induced dipole in rotating-wave-approximation,
\begin{equation}
 \oH_i = - \chi \oE^{+}\oE^{-} \equiv \hbar U_0 \left| f(\ovr) \right|^2 \oa\da \oa, \label{eqn:H_i}
\end{equation}
where we have introduced the coupling constant $U_0 = - \omega_c \chi / 2 \eps_0 V_c$. It has the dimension of a frequency and describes both the optical potential created by a single photon and the shift of the cavity resonance due to the presence of the particle, i.e. the change of index of refraction. Note that if the particle motion is one-dimensional and confined around a position where the mode function can be linearized, $|f(x)|^2 \to x$, one recovers the well known optomechanical interaction Hamiltonian \cite{Kippenberg2008a,Marquardt2009a}.

We describe the interaction in a rotating frame with respect to the pump frequency $\omega_p$ and in a displaced frame with regard to the stationary pump amplitude $\alpha$ such that the driving Hamiltonian for the empty cavity vanishes \cite{Kippenberg2008a,Marquardt2009a}. The mode operator transforms according to $\oa \to \oa + \alpha$. As opposed to previous treatments based on weak field assumptions \cite{Hechenblaikner1998a}, we assume the presence of a \textit{strong pump}, i.e. a significant population of the pumping mode $|\alpha| \gg 1$ and introduce a \textit{first weak coupling condition}:

The coupling to the field should be small compared to the cavity damping rate, $|U_0| \ll \kappa$, so that the presence of the particle introduces only
small fluctuations of the cavity field around the value $\alpha$.

We may therefore neglect the term in the displaced coupling Hamiltonian that is quadratic in the displaced mode operators, and we approximate the total cavity-particle Hamiltonian with
\begin{equation}
 \oH = \hbar \Delta \oa\da \oa + \frac{\ovp^2}{2m_P} + \hbar U_0 \left| f(\ovr) \right|^2 \left( |\alpha|^2 + \alpha^{*} \oa + \alpha \oa\da \right), \label{eqn:Ham0}
\end{equation}
where $\Delta = \omega_c - \omega_p$ is the frequency detuning between the pump and the cavity. The Hamiltonian is the sum of the free cavity energy $\oH_C=\hbar \Delta \oa\da \oa$, the energy of the particle that is moving in the optical potential generated by the stationary pump $\oH_P = \ovp^2/2m_P + \hbar U_0 |\alpha f(\ovr)|^2$, and a linear coupling to the field fluctuations $\oH_I = \hbar U_0 |f(\ovr)|^2 \alpha^{*}\oa + h.c.$. Together with the cavity damping this yields the master equation for the density operator $\rho$ describing the motion of the polarizable particle and the cavity field
\begin{equation}
 \partial_t \rho = -\frac{i}{\hbar} \left[ \oH, \rho \right] + \kappa \left( 2 \oa \rho \oa\da - \oa\da\oa \rho - \rho \oa\da\oa \right) = \left( \cL_P + \cL_I + \cL_C \right) \rho . \label{eqn:ME0}
\end{equation}
It is written in terms of the Liouville superoperators $\cL_{P,I} \rho = -i [\oH_{P,I}, \rho]/\hbar$, and
\begin{eqnarray}
 \cL_C \rho &=& -i\Delta [\oa\da\oa, \rho] + \kappa (2\oa\rho\oa\da - \oa\da\oa \rho - \rho \oa\da\oa) .
\end{eqnarray}
Our goal is now to derive an effective master equation of the particle motion by eliminating the field degrees of freedom. In order to be allowed to adiabatically eliminate the cavity we require the \textit{second, stronger weak coupling condition}:

The effective linear coupling to the field fluctuations is assumed to be small compared to the cavity damping rate, $|U_0 \alpha| \ll \kappa$. This means that any excitation of the cavity on top of the pump amplitude $\alpha$ that is added by a scattering process at the particle will rapidly decay, and $\la \oa\da \oa \ra \ll 1$ in the displaced frame.
As the effective coupling strength is now enhanced by the pump field this represents a much stricter requirement than the first weak coupling condition, and it can be violated in practice by sufficiently strong pump fields, as we will see later.

The second weak coupling condition allows us to use the projection formalism \cite{Stenholm1986a,Cirac1992a,Jaehne2008a} to eliminate the cavity state and derive an effective master equation for the particle. Following this procedure, we introduce the orthogonal superprojector $\cP$ onto the cavity vacuum state (relative to the stationary pump), $\cP \rho = {\rm tr}_C (\rho) \otimes |0 \ra \la 0|$. It fulfills $\cP^2 = \cP$, and its complement is denoted by $\cQ = {\rm id} -  \cP$. The goal is to find a closed evolution equation for $\cP \rho$, which yields the master equation for the reduced particle state $\rho_P = {\rm tr}_C (\rho)$ after tracing out the cavity. Both projectors $\cP, \cQ$ only act on the cavity subspace, and therefore commute with the particle's Liouvillian $\cL_P$. Furthermore, one may easily check that $\cP \cL_C = \cL_C \cP = \cP \cL_I \cP = 0$. This allows us to split the master equation (\ref{eqn:ME0}) into two coupled equations for $\cP\rho$ and $\cQ \rho$ by applying the superprojectors from the left,
\begin{eqnarray}
 \partial_t \cP \rho &=& \cP \cL_P \cP \rho + \cP \cL_I \cQ \rho, \label{eqn:PME}\\
 \partial_t \cQ \rho &=& \cQ \left(\cL_P + \cL_C \right) \cQ \rho + \cQ \cL_I \cP \rho + \cQ \cL_I \cQ \rho \label{eqn:QME}.
\end{eqnarray}
Here, the initial condition is an empty cavity at $t=0$, $\cP \rho(0) = \rho_P (0) \otimes |0 \ra \la 0|$ and $\cQ \rho(0)=0$. The weak coupling condition implies that the contribution of field fluctuations to the state $\cQ \rho$ and to the time evolution $\cL_I$ are suppressed by the perturbation parameter $|U_0 \alpha| / \kappa \ll 1$ on the fast decay time scale $1/\kappa$. We solve the coupled equations up to first order in this parameter to obtain
\begin{equation}
 \partial_t \cP \rho (t) = \cP \cL_P \cP \rho(t) + \int_0^t \diff \tau \, \cP \cL_I \cQ e^{\cL_C\tau} \cQ \cL_I (\tau) \cP \rho (t) \label{eqn:PME2},
\end{equation}
with $\cL_I (\tau) = e^{\cL_P \tau} \cL_I e^{-\cL_P \tau}$ the $\tau$-delayed interaction Liouvillian. We approximate the upper bound of the integral by $t \approx \infty$ for $t \gg 1/\kappa$. This is justified because of the fast field damping in the integrand related to the cavity evolution term $e^{\cL_C \tau}$. Equation (\ref{eqn:PME2}) has to be traced over the cavity to obtain an effective master equation for the reduced particle state $\rho_P(t)$. After insertion of all the Liouvillians, and several steps of straightforward calculation, the effective master equation reads
\begin{equation}
 \partial_t \rho_P = -\frac{i}{\hbar} \left[ \oH_P, \rho_P \right] - U_0^2 |\alpha|^2 \left[ |f(\ovr)|^2, \og \rho_P - \rho_P \og\da \right] \label{eqn:ME1}
\end{equation}
with the operator
\begin{eqnarray}
 \og &=& \int_0^\infty \diff \tau \, e^{-(\kappa + i\Delta) \tau} \left|f \left( \ovr, \ovp; \tau \right) \right|^2. \label{eqn:memop}
\end{eqnarray}
In the following it will be convenient to refer to this object as the \textit{memory operator}. It describes the delayed cavity reaction towards the past trajectory of the particle,
\begin{equation}
 \left|f \left( \ovr, \ovp; \tau \right) \right|^2 = e^{-i \oH_P \tau/\hbar} |f(\ovr)|^2 e^{i \oH_P \tau/\hbar}, \label{eqn:fdelay}
\end{equation}
within the memory time constant $1/\kappa$. We will discuss below that the delay transformation can be approximated by the shearing $|f (\ovr, \ovp; \tau)|^2 \approx \left| f(\ovr - \ovp \tau/m_P ) \right|^2$, if the particle moves sufficiently fast and the influence of the optical potential of the pumped mode is negligible within $1/\kappa$.
Although we started our derivation from a fully Markovian treatment of particle and cavity field, the memory operator renders the effective master equation for the particle non-Lindbladian within the limits of the underlying weak coupling assumptions. This will later result in small negativities of the momentum diffusion coefficient for moving particles, which can be safely neglected in practice.
We will furthermore see that it is the delayed response of the field on the particle motion that is responsible for the optical friction force -- a phenomenon that has been well studied and experimentally observed \cite{Kippenberg2008a,Marquardt2009a} with optomechanical systems. There the one-dimensional motion of a mechanical oscillator is confined at a position where the mode function, that mediates the field coupling, can be linearized. In the more general case of freely moving particles considered here the analoguous friction effect will depend on the particle trajectories through the cavity but can nevertheless be observed.
Compare this e.g.~to polarization gradient cooling \cite{Dalibard1989a}, where it is the delayed response of the internal particle dynamics on the local field polarization which is responsible for the friction force.

\begin{figure}
\centering
\includegraphics[width=8cm]{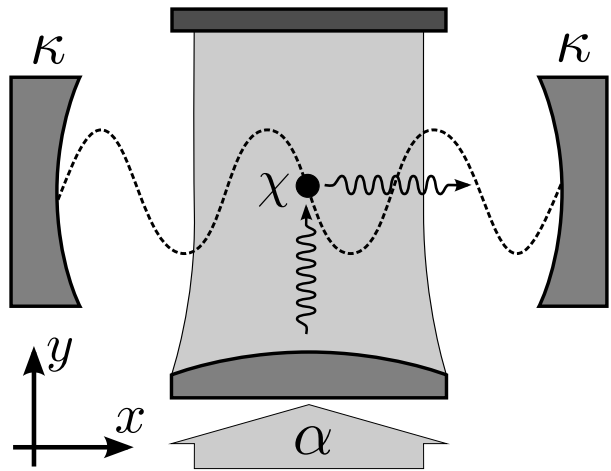}
\caption{\label{fig:cav} Scheme of a two-dimensional mode setup with one $y$-mode strongly pumped to the stationary amplitude $\alpha$, and standing-wave cavity modes orthogonally oriented to the pump along $x$. The latter are damped by the rate $\kappa$ and detuned from the pump. A particle with polarizability $\chi$ experiences dissipative dynamics by scattering photons from the pump to the damped and detuned cavity.}
\end{figure}

A generalization of master equation (\ref{eqn:ME1}) to $N$ particles in $M$ modes is given in \ref{app:generalME}. In the simultaneous presence of $N$ particles in the same pumped mode, our weak coupling criterion reads, however, $ N |U_0 \alpha| \ll \kappa$, and it may be violated by a beam or cloud of weakly coupling highly polarizable particles of sufficient density. Our criterion implies, in particular, that any cross-talk between the particles via field fluctuations in the cavity is small. It therefore also excludes collective correlation effects and self-organization \cite{Domokos2002a,Chan2003a,Black2003a}. At the same time, we do not impose any restriction onto the number $M$ of cavity modes. Using a cavity with many nearly-degenerate modes, such as in a confocal setup \cite{Boyd1961a,Hercher1968a}, one may significantly enhance the control of the particle motion.
Furthermore, the generalized result in \ref{app:generalME} only requires one of the possibly many modes with similar damping and coupling strength to be strongly pumped by $|\alpha| \gg 1$. Our general master equation therefore also includes two-dimensional setups, where the pump light illuminates the particles perpendicularly to the cavity axis \cite{Vuletic2001a}. The master equation of a single particle interacting with $M$ empty modes, and one mode $f_0$ pumped at amplitude $\alpha$, is given by
\begin{equation}
 \partial_t \rho_P = -\frac{i}{\hbar} \left[ \oH_P, \rho_P \right] - \sum_{n=0}^{M} U_{n}^2 |\alpha|^2  \left( \left[ f_{0}^{*} (\ovr) f_{n} (\ovr), \og_{n} \rho_P\right] + h.c. \right). \label{eqn:ME2}
\end{equation}
We here assume the polarizability $\chi$ of the particle to be the same for all modes, i.e. $U_n = - \chi \sqrt{\omega_0 \omega_n / V_0 V_n} / 2 \eps_0 $. The motion of the particle in the conservative potential, given by the Hamiltonian $\oH_P = \ovp^2 / 2m_P + \hbar U_0 \left| \alpha f_0 (\ovr) \right|^2$, is only governed by the optical potential of the pumped mode, while the delayed reaction of the field fluctuations caused by the particle also includes the empty modes through their memory operators
\begin{equation}
 \og_{n} = \int_0^\infty \diff \tau \, e^{-(\kappa_n + i\Delta_n) \tau} \, f_{n}^{*} \left( \ovr ,\ovp ;\tau \right) f_{0} \left( \ovr ,\ovp ;\tau \right). \label{eqn:memop2}
\end{equation}
The product of the mode functions $f_n^{*} f_0$ is evolved back in time in the same manner as in (\ref{eqn:fdelay}). The simple geometry of a particle in an empty Fabry-P\'{e}rot cavity with a transverse pump laser beam is depicted in Figure \ref{fig:cav}. In this case the pumped mode $f_0$ could be a running wave in $y$-direction, while the cavity modes would be one or more standing wave modes in the $x$-direction.

We will discuss in the following section that the cavity memory effect results in friction and diffusion effects which add up with increasing mode number.

\section{The master equation in phase space} \label{sec:phasespace}

In order to study force and diffusion effects, that arise from the delayed cavity feedback on the motion of the particle, it is expedient to translate the underlying master equation into its Wigner-Weyl phase space representation \cite{Wigner1932a,Ozorio1998a,Schleich2001}. Here, the equivalent of the motional quantum state $\rho$ of a particle is given by the Wigner function $w(\vr, \vp)$ of the phase space coordinates $\vr,\vp$. Position and momentum operators acting on the state translate to multiplications and vectorial derivatives by the respective phase space coordinates, accordingly. The details of the translation procedure are given in \ref{app:wigner}.
We treat the particle motion as one-dimensional along the axis of the cooling cavity, i.e. we will resort to scalar phase space coordinates $(x,p)$.
A realistic resonator may often be regarded as one-dimensional if the particle motion perpendicular to the cavity axis is sufficiently well described by a classical trajectory, where the influence of the cavity is negligible. One may then replace the respective position and momentum operators in the master equation by the classical values at time $t$. This yields a parametric time dependence of the mode function, which may be averaged over. A fast directed beam of hot particles crossing the cavity, for example, would experience such an average feedback interaction during the passage time through the cavity. Particles which are trapped in a region overlapping the cavity mode volume, on the other hand, could benefit from a larger average interaction time.

We first translate the master equation (\ref{eqn:ME1}) of a particle in a single pumped mode in one dimension. According to the rules given in \ref{app:wigner}, we arrive at a partial differential equation for the particle Wigner function with an infinite sum of derivatives in $x$ and $p$. A semiclassical approximation that disregards all derivatives higher than second order leads to a Fokker-Planck-type equation
\begin{eqnarray}
 	\fl \partial_t w(x,p) &=& - \partial_x \left(g_x(x,p) w(x,p)\right) - \partial_p \left(g_p(x,p) w(x,p)\right) + \frac{1}{2} \partial_x^2 \left( D_{xx} (x,p) w(x,p)\right) \nonumber \\
	\fl && + \frac{1}{2} \partial_p^2 \left( D_{pp} (x,p) w(x,p)\right) + \partial_x \partial_p \left( D_{xp} (x,p) w(x,p)\right) \label{eqn:FPE}
\end{eqnarray}
which contains the leading order drift term $g_x$, the force $g_p$, and the diffusion terms $D_{xx}$, $D_{xp}$, $D_{pp}$ acting on the particle. We can therefore easily identify the parameters that lead to dissipation and slowing of the polarizable particle. The equation can be split into two parts,
\begin{equation}
 \partial_t w(x,p) = \left[\partial_t w(x,p)\right]_{\rm coh} + \left[\partial_t w(x,p)\right]_{\rm dis}. \label{eqn:ME_Wigner}
\end{equation}
Here, the first part is associated with the coherent evolution under the Hamiltonian $\oH_P$ in (\ref{eqn:ME1}). The second part describes the dissipative cavity reaction. The Hamiltonian evolution corresponds to the quantum-Liouville-Equation in phase space, as discussed in \ref{app:wigner}. It reduces to the classical Liouville equation in the semiclassical limit, that is, it evolves the Wigner function according to the classical equations of motion under the influence of the conservative potential force
\begin{equation}
 g_p^{\rm (coh)} (x,p) = - \hbar U_0 |\alpha|^2 \partial_x \left| f(x) \right|^2 . \label{eqn:conservforce}
\end{equation}
It does not contribute to diffusion. We will therefore omit any further discussion of the coherent part and focus entirely on the dissipative evolution. The crucial step here is to translate the delayed mode function (\ref{eqn:fdelay}) into phase space. By construction, its time derivative fulfills the same quantum-Liouville equation, which we again approximate by the classical Liouville equation. The memory operator (\ref{eqn:memop}) then translates into the memory integral
\begin{equation}
 G(x,p) = \int_0^\infty \diff \tau \, e^{-(\kappa + i\Delta) \tau} \left| f \left( x^{\rm cl}_{-\tau} (x,p) \right) \right|^2, \label{eqn:memint0}
\end{equation}
where the past position $x^{\rm cl}_{-\tau} (x,p)$ at time $-\tau$ is obtained by evolving the phase space coordinates $(x,p)$ back in time along the classical trajectory that is governed by the conservative force (\ref{eqn:conservforce}). This guarantees that the delayed mode function expression fulfills the Liouville equation.

In order to assess the friction and diffusion behaviour analytically, we approximate the past trajectory in the memory integral by the free motion $x^{\rm cl}_{-\tau} (x,p) \approx x - p \tau / m_P$. This corresponds to the above mentioned shearing approximation of Equation (\ref{eqn:fdelay}). We are allowed to do so, if the average momentum change from the conservative force (\ref{eqn:conservforce}) during the memory time $1/\kappa$ is small compared to the momentum $p$ of the particle, that is,
\begin{equation}
 4 \left|\frac{ \omega_r \alpha}{k v} \right| \, \left|\frac{U_0 \alpha}{\kappa} \right| \ll 1 . \label{eqn:shearingcond}
\end{equation}
Here, we used the fact that the derivative of the mode function in (\ref{eqn:conservforce}) gives a factor of the order of the wave number, $\partial_x f(x) \sim k f(x)$. We denote the Doppler shift of the moving particle by $k v = k p /m_P$. The {\em recoil frequency} $\omega_r = \hbar k^2 / 2 m_P$ is the kinetic energy transferred to the particle by a single scattered photon, which is a small quantity for the large masses considered in this article. In total, the above condition is violated when either the particle is so slow or the pump amplitude $\alpha$ is so strong that the optical potential of the pumped mode starts to act as a trap for the particle. We do not consider this case here, and we approximate the memory integral from here on as
\begin{equation}
 G(x,p) = \int_0^\infty \diff \tau \, e^{-(\kappa + i\Delta) \tau} \left| f \left( x - \frac{p \tau}{m_P} \right) \right|^2. \label{eqn:memint1}
\end{equation}

Putting everything together, we arrive at the force and diffusion terms of the dissipative evolution
\begin{eqnarray}
 g_p^{\rm (dis)} (x,p) &=& \hbar U_0^2 |\alpha|^2 \re \left[ 2i G(x,p) \partial_x \left| f(x) \right|^2 \right.  \nonumber \\
 && \left. - \hbar \left( \partial_p G(x,p) \right) \partial_x^2 \left| f(x) \right|^2 \right] \label{eqn:gp1}\\
 D_{pp}^{\rm (dis)} (x,p) &=& 2 \hbar^2 U_0^2 |\alpha|^2 \re \left[ \partial_x G(x,p) \right] \partial_x \left| f(x) \right|^2 \label{eqn:Dpp1} \\
 D_{xp}^{\rm (dis)} (x,p) &=& -\hbar^2 U_0^2 |\alpha|^2 \re \left[\partial_p G(x,p) \right] \partial_x \left| f(x) \right|^2 \label{eqn:Dxp1}
\end{eqnarray}
There is no contribution to the coefficients $g_x$ and $D_{xx}$ in the Fokker-Planck equation (\ref{eqn:FPE}). The second line in the force term (\ref{eqn:gp1}) represents the second order correction in $\hbar$ to the dissipative force, while the first line is of order $\Order{\hbar^1}$. Again, we can generalize the procedure to $N$ particles in $M$ cavity modes to obtain the dissipative terms as given in \ref{app:generalFPE}.

The Fokker-Planck coefficients (\ref{eqn:gp1}) and (\ref{eqn:Dpp1}) are the basis for studying the motional damping of the particle due to the delayed reaction of the cavity. A closer look at the velocity dependence of the dissipative force (\ref{eqn:gp1}) reveals whether the particle is slowed or accelerated by the cavity. At low particle velocities $v = p/m_P$ the force can be expanded to first order,
\begin{equation}
 g_p^{\rm (dis)} (x,p) = g_p^{\rm (dis)} (x,0) + \beta(x) p + \Order{p^2} \label{eqn:lowv_friction}
\end{equation}
where the second term represents the friction force acting on the particle. Motional damping is indicated by a negative sign of its friction coefficient $\beta(x) = \partial_p g_p^{\rm (dis)} (x,p\to 0)$, which is
\begin{eqnarray}
 \fl \beta(x) = - \frac{4 \hbar \kappa U_0^2 |\alpha|^2}{m_P (\kappa^2 + \Delta^2)^2} \left( \Delta \left( \partial_x \left| f(x) \right|^2 \right)^2 +  \frac{\hbar}{2m_P} \frac{\kappa^2 - \Delta^2}{\kappa^2 + \Delta^2} \left( \partial_x^2 \left| f(x) \right|^2 \right)^2 \right). \label{eqn:beta1}
\end{eqnarray}
If $\beta(x) < 0$ the coefficient can be seen as a momentum damping rate in units of Hz. The derivative of the mode function $f(x)$ gives the factor $\partial_x f(x) \sim k f(x)$. This leads to the criterion $|k v| < \kappa$ for the applicability of the low-velocity expansion (\ref{eqn:lowv_friction}): The dissipative force can no longer be described in terms of the friction coefficient (\ref{eqn:beta1}) if the particle travels more than the distance $1/k$ within the cavity reaction and decay time $1/\kappa$.
The first and the second line in (\ref{eqn:beta1}) correspond to the friction in first and second order of the recoil frequency $\omega_r$, respectively.

It is interesting to note that the sign of both contributions to the friction is fixed by the cavity-pump detuning $\Delta$, irrespective of the particular mode function. In particular, red-detuned pump frequencies, i.e. $\Delta >0$, yield a momentum damping in first order of the recoil frequency and our result for the friction then agrees with the results in \cite{Hechenblaikner1998a,Domokos2003a}, which were obtained in a purely semiclassical approach with Langevin equations of motion.

This represents the intuitive picture of enhanced Doppler-shifted backward-scattering of a pump photon, transferring kinetic energy of the order of $\hbar \omega_r$ to a blue-detuned and rapidly damped cavity photon. The effect is strongest at the steepest slope of $|f(x)|^2$, and it vanishes at the extrema of the optical potential. Consequently, the friction force is significantly suppressed when the optical potential is deep enough to trap the particle at a potential minimum.

The optimal detuning maximizing the magnitude of the first order term in (\ref{eqn:beta1}) is $\Delta = \kappa / \sqrt{3}$. Putting everything together, the position-averaged friction then scales as
\begin{equation}
 \overline{\beta} \sim - \frac{3 \sqrt{3} }{2} \left| \frac{U_0 \alpha}{\kappa} \right|^2 \omega_r + \Order{\frac{\omega_r^2}{\kappa}}. \label{eqn:beta_av1}
\end{equation}
We find that the momentum damping is a second-order effect in the weak-coupling parameter $U_0 |\alpha| / \kappa$. Note that a very weak polarizability contained in $U_0$ can be largely compensated by a sufficiently strong field $\alpha$. For too strong $\alpha$ the condition of weak coupling is not fullfilled anymore which, however, does not necessarily mean that no cooling occurs. The friction is in addition proportional to the recoil frequency, which decreases with growing mass of the particle. In order to achieve a significant slowing of the particle it has to interact with the pumped mode over a time of the order $1 / \overline{\beta}$. A similar calculation shows that the position-averaged momentum diffusion coefficient scales as
\begin{equation}
 \overline{D_{pp}} \sim \frac{3 \hbar m_P}{2} \left| \frac{U_0 \alpha}{\kappa} \right|^2 \kappa \omega_r \label{eqn:Dpp_av1}
\end{equation}
at low particle velocity.

The discussion of the friction term in (\ref{eqn:beta1}), that is of second order in $\omega_r$, is omitted here because it is negligible everywhere except for the extrema of the optical potential and on resonance $\Delta = 0$ or $|\Delta| \gg \kappa$, where the first order term vanishes. At a detuning larger than the cavity linewidth, $|\Delta| > \kappa$, the second order term anti-damps the momentum of the particle. Higher order contributions to the force could be obtained beyond the semiclassical approximation in the phase space translation (\ref{eqn:ME_Wigner}) of the master equation. Such contributions are suppressed by increasing powers of $\omega_r / \kappa$. This justifies, \textit{a posteriori}, the semiclassical approximation for sufficiently large particles, where $\omega_r / \kappa \ll 1$.

Realistic friction rates can be estimated from the characteristic coupling parameters given in Table~\ref{tab:mol} for a number of large polarizable particles. We consider a Gaussian mode with a typical mode volume $V=0.1\,$mm$^3$ and a damping rate $\kappa = 1\,$MHz, that is pumped by an IR laser of wavelength $\lambda = 1.5\,\mu$m. A laser with a power of $P_L = 0.25\,$W would pump this mode to a stationary photon number $|\alpha|^2 = P_L / 2 \kappa \hbar \omega_p \approx 10^{12}$. At such a power the
single-particle weak coupling condition $|U_0 \alpha| \ll \kappa$ would still be fulfilled also for the heavy particle examples in the table. The average friction expression (\ref{eqn:beta_av1}) thus holds and yields the velocity damping rates $\overline{\beta} = 38~$Hz, $3~$Hz, and $0.8~$Hz for the lithium, the silica, and the gold cluster in the table, respectively. Although the polarizabilities of the three cluster types are roughly the same, the lithium cluster profits here from its comparably small mass. Compared to the metal clusters, both the C$_{60}$ fullerene and a helium nanodroplet of 1000 atoms have a much smaller damping rate of about 100~mHz, whereas a single lithium atom reaches 800~mHz because of its light weight.
Note that the laser and cavity parameters used here could lead to a violation of the weak coupling condition by the simultaneous presence of already a few of the heavy clusters in the table or of about $1000$ fullerene particles, for instance.

If the weak coupling condition is violated, the friction and diffusion terms derived here cease to be valid. However, complementary semiclassical approaches to this problem and numerical simulations in the strong coupling regime \cite{Hechenblaikner1998a,Domokos2003a} suggest that the friction should generally scale further with the coupling strength beyond the weak coupling limit. In general, the effective strong coupling regime $|U_0 \alpha| > \kappa$ can be achieved in the experiment with large, and thus highly polarizable, atomic clusters in connection with a strong pump laser power. On the other hand, the counteracting decrease of the recoil frequency with the mass influences both the conservative and the dissipative forces acting on the particle, as well.

There are additional momentum diffusion effects due to the absorption and free-space scattering of the laser light to be considered, which accompany the diffusion caused by the cavity alone. If they are comparable to or larger than the cavity diffusion effect they will significantly influence the cooling limit of an ensemble of particles, i.e. the final kinetic energy reached by the friction process. The additional diffusion effects will be discussed in Section \ref{sec:MEabsscatt}. A simple damping-diffusion time evolution model for the velocity \cite{Cohen1992,Hechenblaikner1998a} may then be applied to estimate the final kinetic energy where the friction effect and the momentum diffusion due to the cavity compensate each other,
\begin{equation}
\lla \frac{p^2}{2m_P} \rra_\infty \sim - \frac{D_{pp}^{\rm (dis)} (x,p\to 0)}{2 m_P \beta(x)} \approx \frac{\hbar}{4} \left( \Delta + \frac{\kappa^2}{\Delta} \right). \label{eqn:limit1}
\end{equation}
Here, we used the fact that the momentum diffusion coefficient (\ref{eqn:Dpp1}) exhibits the same functional dependence of the mode function as the first order friction coefficient in (\ref{eqn:beta1}) in the limit of small particle velocity. The expression (\ref{eqn:limit1}) depends on the detuning and ceases to be valid in the extreme cases $\Delta=0$, $\Delta \gg \kappa$, where the first order friction vanishes. In case of the above optimal detuning the final energy would be $\hbar \kappa / 2 \sqrt{3}$. This recovers the result of the single-particle cavity cooling limit and agrees with the results of wavefunction simulations in \cite{Horak1997a}. For an experiment it suggests the use of very good cavities to obtain faster cooling by larger field amplitudes together with lower temperatures $\kappa$.

\begin{table*}[htb]
  \caption{Light coupling parameter table for different polarizable particles and IR laser light with a wavelength $\lambda = 1.5 \, \mu$m. The polarizability $\chi$ and the absorption cross section $\sigma_a$ at this wavelength are listed for particles of different mass $m$, with a focus on heavy-sized clusters of gold, lithium and fused-silica. Their optical properties are taken from \cite{Palik1985,Gorodetsky2000a}. For the lithium atom and the superfluid He$_{1000}$ droplet the polarizability is estimated by its static value per atom \cite{Miffre2006a,Lach2004a}. Both He and Li are transparent to IR light and we omit their absorption here. For the C$_{60}$-fullerene we extract both $\chi$ and $\sigma_a$ from the spectra in \cite{Dresselhaus1998a}. The derived values of the recoil frequency $\omega_r$, the coupling constant $U_0$, the photon absorption rate $2\gamma_a$, and the Rayleigh scattering rate $2\gamma_s$ are given in units of MHz, which corresponds to typical, experimentally achievable cavity damping rates $\kappa$. The cavity mode volume is set to $V=0.1\,$mm$^3$. \label{tab:mol}}
\begin{footnotesize}
\begin{tabular}{cccccccc}
\hline
Particle & $m$ [amu] & $\chi$ [\AA$^3 \times 4\pi \varepsilon_0$] & $\sigma_a$ [\AA$^2$] & $\omega_r$ [MHz] & $|U_0|$ [MHz] & $2\gamma_a$ [MHz] & $2\gamma_s$ [MHz]\\
\hline
Li & $7$ & $24$ &  & $8.0 \cdot 10^{-2}$ & $1.9 \cdot 10^{-9}$ &  & $8.9 \cdot 10^{-18}$ \\
C$_{60}$ & $720$ & $83$ & $\sim 10^{-4}$ & $7.7 \cdot 10^{-4}$ & $6.5 \cdot 10^{-9}$ & $\sim 10^{-12}$ & $1.0 \cdot 10^{-16}$ \\
He$_{1000}$ & $4000$ & $200$ &  & $1.4 \cdot 10^{-4}$ & $1.6 \cdot 10^{-8}$ &  & $6.2 \cdot 10^{-16}$ \\
Li$_{1000}$ & $7000$ & $5501$ & $2.6 \cdot 10^{-1}$ & $8.0 \cdot 10^{-5}$ & $4.3 \cdot 10^{-7}$ & $1.5 \cdot 10^{-8}$ & $4.7 \cdot 10^{-13}$ \\
(SiO$_2$)$_{1000}$ & $60000$ & $2901$ & $6.1 \cdot 10^{-11}$ & $2.0 \cdot 10^{-5}$ & $2.3 \cdot 10^{-7}$ & $3.7 \cdot 10^{-18}$ & $1.3 \cdot 10^{-13}$ \\
Au$_{1000}$ & $197000$ & $4180$ & $8.2 \cdot 10^{-2}$ & $2.8 \cdot 10^{-6}$ & $3.3 \cdot 10^{-7}$ & $4.8 \cdot 10^{-9}$ & $2.7 \cdot 10^{-13}$ \\
\hline
\end{tabular}
\end{footnotesize}
\end{table*}

\subsection{Fabry-P\'{e}rot cavity}

\begin{figure}
\centering
\includegraphics[width=10cm]{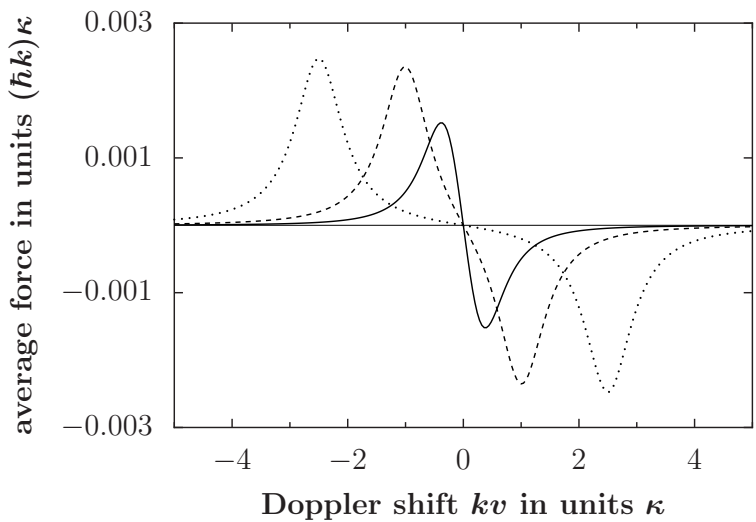}
\caption{\label{fig:FPvel} Position-averaged dissipation force (\ref{eqn:gpFP}) plotted versus the particle velocity for different cavity-pump detunings, using an effective coupling strength $|U_0 \alpha| = 0.1 \kappa$. The velocity is expressed in terms of the Doppler shift $kv$ in units of the mode damping $\kappa$, and the force is plotted in units of the photon momentum transfer rate $(\hbar k) \kappa$. The solid line represents the detuning $\Delta = \kappa / \sqrt{3}$, which gives the the optimal friction coefficient (\ref{eqn:beta_av1}), that is, the steepest slope around $kv=0$. The dashed and the dotted line represent the higher detuning $\Delta = 2\kappa$ and $\Delta = 5 \kappa$, respectively. Although they exhbit a weaker damping of small velocities, their damping range of velocities grows.}
\end{figure}

For the instructive example of a one-dimensional Fabry-P\'{e}rot cavity with mode function $f(x) = \cos kx$ the memory integral (\ref{eqn:memint1}) and thus the dissipative force and diffusion terms can be calculated explicitly for any given velocity. Focussing on their dependence of the particle velocity, we take the average over the position coordinate and obtain
\begin{eqnarray}
 \overline{g_p^{\rm (dis)}} (m_P v) &=& - 2\hbar k \frac{\kappa |U_0 \alpha|^2 \Delta kv }{\left| \nu^2 + (2kv)^2 \right|^2} \label{eqn:gpFP}\\
 \overline{D_{pp}^{\rm (dis)}} (m_P v) &=& (\hbar k)^2 \frac{\kappa |U_0 \alpha|^2 \left( |\nu|^2 + (2kv)^2 \right) }{\left| \nu^2 + (2kv)^2 \right|^2} \label{eqn:DppFP}
\end{eqnarray}
with the complex parameter $\nu = \kappa + i\Delta$. The force term is given in leading order of the recoil frequency $\omega_r$. Both terms scale identically with the effective coupling to the cavity field $U_0 |\alpha|$. Moreover, they vanish in the \textit{Doppler resolution limit} $kv \ll \kappa,\Delta$, where the delayed cavity reaction is too slow to affect the particle's motion.
The velocity dependence of the position-averaged dissipative force is plotted in Fig.~\ref{fig:FPvel} for an effective coupling $|U_0 \alpha| = 0.1 \kappa$ and different detunings $\Delta$.
One clearly sees the linear frictional behaviour of the force term in a small window around zero velocity, where the detuning $\Delta = \kappa /\sqrt{3}$ gives the optimal slope. On the other hand, the range of velocities that are damped by the dissipative force term increases with growing detuning $\Delta$. In the limit of large Doppler shifts $|kv| \gg \kappa$ the damping effect is maximized for $\Delta \approx |k v|$. Fast particles are therefore best addressed with a detuning far off the cavity resonance.
Note that in order to combine large capture ranges and strong friction for low velocities one could think of a time modulation of the detuning or the simultaneous use of various cavity modes operated at different detunings.

Note that our approximation of the memory integral (\ref{eqn:memint1}) formally fails in close vicinity of the point $kv=0$, where the condition (\ref{eqn:shearingcond}) is violated. Although, strictly speaking, the discussed friction force is therefore not valid for such small velocities, we expect the error to be negligible for typical pump amplitudes that do not by themselves already create an optical potential that is strong enough to dominate the particle motion.

\begin{figure}
\centering
\includegraphics[width=10cm]{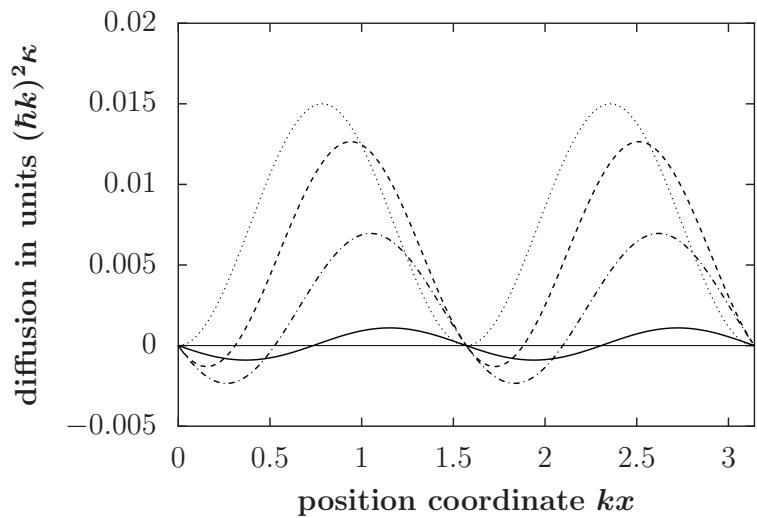}
\caption{\label{fig:FPdiff} Momentum diffusion coefficient (\ref{eqn:DppFP2}) plotted as a function of position for several fixed velocities. The particle's position $x$ is expressed in terms of the dimensionless parameter $kx$ in the cavity field, and the diffusion coefficient is given in units of $(\hbar k)^2 \kappa$. The curves correspond to the momentum diffusion expressions for the fixed Doppler shifts $kv=0$ (dotted line), $kv=0.5\kappa$ (dashed line), $kv=\kappa$ (dash-dotted line), and $kv=5\kappa$ (solid line). The negative part of the diffusion coefficient grows with increasing modulus of the velocity, until it practically compensates the positive part. Negative velocities give the same results due to symmetry. Only the diffusion for $kv=0$ is non-negative everywhere.}
\end{figure}

We also want to point out that the positivity of the average momentum diffusion (\ref{eqn:DppFP}) does not guarantee the positivity of the diffusion coefficient at every point $(x,p)$ in phase space. This is in fact not the case in our semiclassical approximation, as we will see below. The positive semidefiniteness of the diffusion matrix $(D_{ij})$ is required to interpret the semiclassical Wigner function evolution equation that is derived from our master equation as a classical Fokker-Planck-Equation \cite{Risken1989}. Only then does it make sense to translate the phase space equation into an equivalent classical Langevin equation for the position and momentum coordinates. However, in the Fabry-P\'{e}rot case considered here, the momentum diffusion coefficient (\ref{eqn:Dpp1}) reads
\begin{eqnarray}
D_{pp}^{\rm (dis)} (x,m_P v) &=& \frac{2 (\hbar k)^2 |U_0 \alpha|^2}{\left| \nu^2 + (2kv)^2 \right|^2} \left[ \kappa \left( |\nu|^2 + (2kv)^2 \right) \sin^2 (2kx) \right.\nonumber \\
&&- \left. 2kv \left( \re (\nu^2) + (2kv)^2 \right) \sin (2kx) \cos (2kx) \right] \label{eqn:DppFP2}
\end{eqnarray}
As demonstrated in Fig.~\ref{fig:FPdiff} it is not positive everywhere for nonzero velocities.
The diffusion coefficient takes on small negative values close to the minima of the optical potential, but it is still dominantly positive for slowly moving particles, $|kv| \lesssim \kappa$. In the case of faster particles, where the negative and positive regions almost compensate each other, the momentum diffusion effect vanishes rapidly.
We attribute the negativity to the non-Markovian nature of the underlying master equation (\ref{eqn:ME1}). The memory effect of the cavity, and in particular the accompanied correlation between the particle motion and the cavity field fluctuations \cite{Domokos2001a}, impede a completely positive Lindblad form \cite{Barnett2001a,Salo2006a,Piilo2009a}. The effect is however small and negligible in the practical considerations on the motional damping of large polarizable particles discussed in the following.

\subsection{Confocal cavity}

As mentioned in Section \ref{sec:me}, the effective friction force should increase in the presence of $M$ near degenerate empty cavity modes $f_1, \ldots, f_M$ in addition to the pumped mode $f_0$. Again, this can be best studied after translating the corresponding master equation (\ref{eqn:ME2}) into phase space; this is a special case of the general formalism in \ref{app:generalFPE}. Disregarding the Hamiltonian evolution and considering only one dimension of motion, the semiclassical dissipative force and diffusion coefficients read as
\begin{eqnarray}
 g_p^{\rm (dis)} (x,p) &=& \hbar \sum_{m=0}^M |U_m \alpha|^2 \re \left[ 2i G_m(x,p) \partial_x f_0^{*} (x) f_m (x) \right.  \nonumber \\
 && \left. - \hbar \left( \partial_p G_m(x,p) \right) \partial_x^2 f_0^{*}(x) f_m (x) \right], \label{eqn:gp2}\\
 D_{pp}^{\rm (dis)} (x,p) &=& 2 \hbar^2 \sum_{m=0}^M |U_m \alpha|^2 \re \left[ \partial_x G_m(x,p) \partial_x f_0^{*}(x) f_m (x) \right], \label{eqn:Dpp2} \\
 D_{xp}^{\rm (dis)} (x,p) &=& -\hbar^2 \sum_{m=0}^M |U_m \alpha|^2 \re \left[\partial_p G_m(x,p) \partial_x f_0^{*}(x) f_m (x) \right], \label{eqn:Dxp2}
\end{eqnarray}
where we introduced the memory integral for each mode in the same free-motion-approximation as before
\begin{equation}
 G_m(x,p) = \int_0^\infty \diff \tau \, e^{-(\kappa_m + i\Delta_m) \tau} f_m^{*} \left( x - \frac{p \tau}{m_P} \right) f_0 \left( x - \frac{p \tau}{m_P} \right). \label{eqn:memint2}
\end{equation}
Both the force and the diffusion coefficient are sums over terms similar to the single-mode case (\ref{eqn:gp1})-(\ref{eqn:Dxp1}). The only difference is that the modulus square of a single-mode function is replaced by the cross-products $f_0^{*} f_m$ and the complex conjugates. In the limit of small velocities the friction coefficient now reads
\begin{eqnarray}
 \beta(x) &=& - \sum_{m=0}^M \frac{4 \hbar \kappa_m |U_m \alpha|^2}{m_P (\kappa_m^2 + \Delta_m^2)^2} \left( \Delta_m \left| \partial_x \left( f_0^{*} (x) f_m (x) \right) \right|^2 \right. \nonumber \\
&&+ \left. \frac{\hbar}{2m_P} \frac{\kappa_m^2 - \Delta_m^2}{\kappa_m^2 + \Delta_m^2} \left| \partial_x^2 \left( f_0^{*} (x) f_m (x) \right) \right|^2 \right). \label{eqn:beta2}
\end{eqnarray}
Again, the sign of the low-velocity friction is independent of the mode functions. It is solely determined by the signs of the mode detunings $\Delta_m$ with respect to the pump frequency.

The multi-mode friction force can be illustrated with the practical example of a confocal standing wave cavity. In order to apply our one-dimensional description here, we neglect all dynamical effects due to the transverse mode profile, and we treat the transverse coordinates $y,z$ as parameters to arrive at effectively one-dimensional modes along the cavity axis $x$. In the end, we take the average of the resulting friction coefficient (\ref{eqn:beta2}) over the off-axis coordinates $y,z$. By setting the averaging area much larger than the cavity mode extensions, this approximation is relevant for experiments where a directed molecular beam crosses the cavity to obtain transverse motional cooling. The particle ensemble is assumed to be sufficiently dilute so that only one particle passes the cavity at any given time. In a typical case, when the mode waists are much larger than the laser wavelength, the transverse dissipation and dipole forces in $y,z$ can be safely neglected.

\begin{figure}
\centering
\includegraphics[width=10cm]{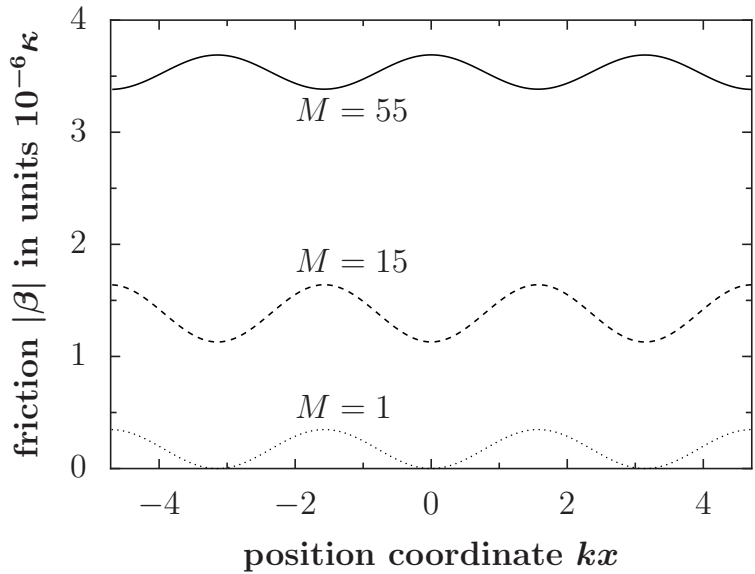}
\caption{\label{fig:confocal_beta1} Plot of the modulus of the multi-mode friction coefficient (\ref{eqn:beta2}) of a confocal resonator versus the position $kx$ along the cavity axis. The transverse coordinates $y$ and $z$ are averaged over a large area with respect to the waist of the contributing cavity modes. The pump mode is given by a Gaussian running wave mode in $y$-direction. See the text for numerical details. The dotted curve represents the friction related to the coupling of the pump mode to a single fundamental mode of the cavity only, whereas the pump mode couples to $M=15$ and $M=55$ degenerate modes (including the fundamental mode) in the dashed and in the solid lined plot, respectively. The friction coefficient is negative in all cases, and it is plotted in units of $10^{-6}$ times the linewidth of the cavity modes $\kappa$.}
\end{figure}

We consider a confocal cavity with two circularly symmetric mirrors of sufficiently large aperture, where the mode functions are well described by Laguerre-Gaussian standing-wave modes \cite{Boyd1961a,Hodgson2005}. They are characterized by a longitudinal mode index $n\in \N$ and two transverse mode indices $m,\ell \in \N_0$ that yield the resonance wavenumbers
\begin{equation}
 k_{n,m,l} = \frac{\pi}{d} \left( n + \frac{2m + \ell + 1}{2} \right).
\end{equation}
The distance $d$ of the mirrors equals their radii of curvature. Hence, for a given fundamental mode $(n_0,0,0)$ there exists an infinite number of degenerate higher-order transverse modes $(n < n_0,m,\ell)$ at the same resonance frequency. In practice, the extended transverse profiles of higher order modes, however, limit the degeneracy as the mode damping $\kappa_{n,m,\ell}$ increases with growing $m,\ell$. This can be expressed in terms of the effective mode waist \cite{Hodgson2005}
\begin{equation}
 w_{n,m,\ell} = w_{0} \sqrt{2m + \ell + 1}. \label{eqn:effwaist}
\end{equation}
It grows with respect to the waist $w_0$ of the corresponding fundamental mode $(n_0,0,0)$. The maximally achievable waist without a significant loss of cavity finesse then serves as a rough estimate of the number $M$ of degenerate modes: If a confocal cavity for example tolerates an effective waist without considerable losses, that is $a$ times larger than the waist of the pumped fundamental mode, then the transverse mode index expression $2m+\ell+1$ may have values up to $a^2$, where both $m$ and $\ell$ can vary independently. This results in a degeneracy of the order of $a^4/2$ modes which contribute to the average friction.
These modes include both cosine and sine standing wave modes for $n$ even and odd, respectively. The advantage of this is demonstrated in Fig.~\ref{fig:confocal_beta1}, where the one-dimensional friction coefficient (\ref{eqn:beta2}) is plotted as a function of $x$-position around the center of a confocal standing-wave resonator for different contributing mode numbers. We consider the fundamental mode $n_0 = 2 \cdot 10^4$ of the confocal cavity of size $d = 10\,$mm to be coupled to a Gaussian running wave mode $f_0$ in $y$-direction with the same waist $w_0$, following the scheme in Fig.~\ref{fig:cav}. The $y$-mode is strongly pumped by an IR laser detuned with respect to the cavity to $\Delta = \kappa/\sqrt{3}$, and the mode wavelength is given by $\lambda = 2d/(n_0+1/2) = 1 \, \mu$m. We assume that all degenerate modes share the same common linewidth $\kappa$. The pump amplitude $\alpha$ is fixed to a value that corresponds to the effective coupling $U_{n_0}\alpha = 0.1 \kappa$ between the pumped Gaussian mode and the fundamental cavity mode.
We set the recoil frequency of the particle to $\omega_r = 0.001 \kappa$ in this case, which corresponds to a mass $m_P = 1000\,$amu and a cavity linewidth $\kappa = 1.25\,$MHz.
The $x$-derivative of the running-wave mode $f_0$ may be neglected in the multi-mode friction formula (\ref{eqn:beta2}), i.e. the friction is only caused by the coherent scattering of pump photons into the degenerate confocal cavity modes. The transverse coordinates $y$ and $z$ are each averaged out numerically over a distance of 4 times the fundamental waist off the central cavity axis.
Fig.~\ref{fig:confocal_beta1} shows that the average friction not only increases with growing mode number, as expected, but it also does not oscillate to zero anymore. The latter is due to the equal presence of both sine and cosine standing wave modes.

\begin{figure}
\centering
\includegraphics[width=10cm]{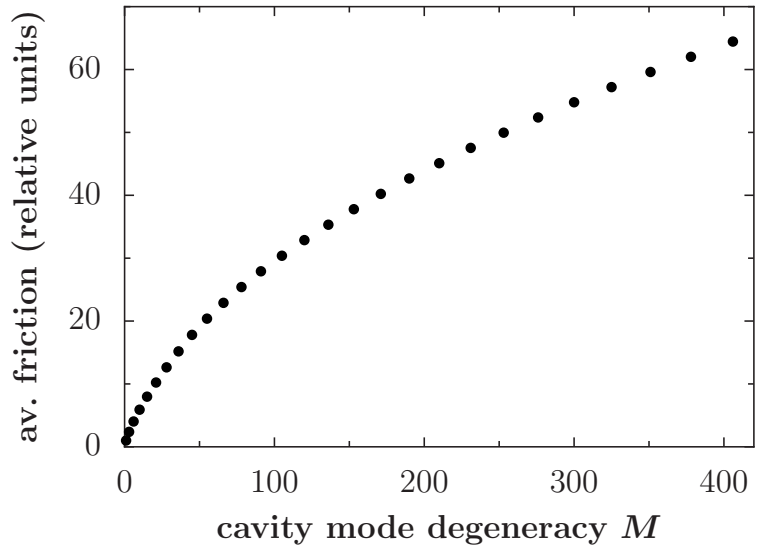}
\caption{\label{fig:confocal_beta2} Modulus of the position-averaged friction coefficient of Fig.~\ref{fig:confocal_beta1} plotted as a function of the number $M$ of contributing degenerate modes in the confocal cavity. The first data point represents the friction related to the coupling of the Gaussian pump mode to the single fundamental mode $(n_0=2 \cdot 10^4,0,0)$ of the cavity. It corresponds to the average of the dotted curve in Fig.~\ref{fig:confocal_beta1}. All data points in this plot are given relative to this value. The rightmost data point contains the friction contribution of $M=406$ degenerate modes including the fundamental one. }
\end{figure}

In practice, all degenerate high-finesse modes, which are supported by the confocal resonator, contribute to the friction. The dependence of the friction coefficient (\ref{eqn:beta2}) on the supported degeneracy is depicted in Fig.~\ref{fig:confocal_beta2}. Here, the one-dimensional friction of Fig.~\ref{fig:confocal_beta1} is in addition averaged over $x$ and plotted as a function of the number of contributing degenerate modes. The mean friction experienced by each member of the particle ensemble crossing the cavity increases with each higher-order mode that is supported by the cavity.
However, Fig.~\ref{fig:confocal_beta2} does not indicate a linear growth of the average friction with the degeneracy. This can be explained with the multi-mode friction formula (\ref{eqn:beta2}). Each empty cavity mode $f_m$ enters there in a product with the Gaussian pump mode $f_0$, which has a fixed spatial extension given by its waist $w_0$. Then again, the transverse profile of higher order modes stretches over a larger area off the cavity axis as their effective waist (\ref{eqn:effwaist}) grows. Consequently, their overlap with the pump mode descreases, and the multi-mode friction coefficient saturates for a sufficiently large degeneracy.

\section{Light absorption and scattering} \label{sec:MEabsscatt}

So far we have assumed the particle to have a real, scalar and positive linear response $\chi$ to the optical cavity field, i.e. to be an ideal high-field-seeker with no photonic loss channels such as absorption or scattering into free-space.
This has led to the derivation of an effective master equation in Section \ref{sec:me} and yielded the analytical expressions for friction and diffusion in Section \ref{sec:phasespace}.
In practice this condition holds fairly well for two-level-atoms in optical fields far red-detuned from their internal resonance
\cite{Vukics2005a}. It is also a good description of large molecules or clusters with sufficiently low absorption, where the large number of internal degrees of freedom in general does not allow us to formulate a simple few-state coupling model. Nevertheless, many theoretical approaches towards dissipative cavity cooling of molecules \cite{Lev2008a,Deachapunya2007a,Lan2010a} start from an effective two-level-description of the molecule's internal dynamics, which is then adiabatically eliminated in the off-resonant limit to arrive at a real polarizability again. In such a formulation one can easily incorporate spontaneous emission of photons into the vacuum, which yields a free-space Rayleigh scattering rate $\gamma_r$ in the off-resonant limit.

In the following we discuss the influence of scattering and absorption at large particles in terms of experimentally accessible scattering and absorption cross sections. We introduce additional Lindblad terms in our master equations to account for these effects, and we translate them into phase space in order to reveal their influence on dissipation and momentum diffusion.
The dominant effect is caused by the pump photons and we neglect the influence of field fluctuations or thermal photons.

The derivation of the additional Lindblad terms is given in \ref{app:abs}, where we apply the same weak coupling model as used in Sec.~\ref{sec:me} to couple the pump field to internal degrees of freedom of the particle (to describe absorption) and to free-space vacuum modes of light (to describe Rayleigh scattering). The weakness of the coupling in these cases is a reasonable assumption for the large particles and the off-resonant pump light considered here. This is illustrated by the examples in Table~\ref{tab:mol}, where the the single-photon absorption and scattering rates $\gamma_a, \gamma_s$ are already significantly smaller than the single-photon coupling $U_0$ to the cavity mode.

\subsection{Absorption}

In order to achieve dissipation and eventually cooling of a molecular ensemble we have to avoid absorption which would lead to heating. Dipole allowed transitions in many organic molecules as well as plasmon resonances in metal clusters are typically situated in the UV-VIS range. This makes it advisable to pump the cavity with near-infrared light, i.e. the extreme tail of the absorption spectrum.
Each particle may then be regarded as a thermal bath of harmonic oscillators with a linear and approximately constant coupling to the external cavity field modes (see \ref{app:abs}). When the particles remain internally cold during their residence time in the cavity, we are allowed to trace over the internal oscillators to obtain the Lindblad term for the pumped cavity mode
\begin{equation}
 \left[ \partial_t \rho_P \right]_{\rm abs} = \gamma_a |\alpha|^2 \left( 2 f (\ovr) \rho_P f^{*} (\ovr) - \left\{ \rho_P , \left| f (\ovr) \right|^2 \right\}  \right). \label{eqn:absLind}
\end{equation}
This term has to be added to the master equations (\ref{eqn:ME1}) and (\ref{eqn:ME2}) in order to account for the momentum diffusion that is caused by the recoil of absorbed pump photons. The absorption rate per photon $\gamma_a$ can be expressed in terms of the cross section $\sigma_a$ as $2\gamma_a = c \sigma_a / V$.

\subsection{Rayleigh scattering}

The elastic scattering of pump photons into free-space can be accounted for by the standard Lindblad term
\begin{eqnarray}
 \left[ \partial_t \rho_P \right]_{\rm sca} &=& 2 \gamma_s |\alpha|^2 \int_{|\vu |=1} \diff^2 \vu \, N(\vu)  \ok_{\vu} \rho_P \ok\da_{\vu} - \gamma_s |\alpha|^2 \left\{ \rho_P , \left| f (\ovr) \right|^2 \right\} \label{eqn:scattLind}\\
 \ok_{\vu} &=& e^{-i\omega_p \vu \cdot \ovr / c} f (\ovr)
\end{eqnarray}
As shown in \ref{app:abs}, this expression is obtained by modelling the space outside the cavity as a band of vacuum modes with arbitrary detuning in a master equation of the form (\ref{eqn:ME2}). The coupling parameters are then replaced by the Rayleigh scattering rate $\gamma_s$ and the angular scattering distribution $N(\vu)$.
Any Doppler shift related to the scattering by moving particles can be neglected since we are only interested in the additional momentum diffusion that is caused by the random recoil of the scattered photons.
In a one dimensional situation, where only the particle motion along the cavity axis $x$ is taken into account, one can resort to an effective mode function $f(x)$. In this case the scattering Lindblad operators are replaced by
\begin{equation}
 \ok_{\vu} = e^{-i\omega_p u_x \ox / c} f (\ox),
\end{equation}
with $u_x$ the component of the vector $\vu$ along the cavity axis.
Recall our sans-serif notation for the position operator $\ox$ that distinguishes it from the scalar value $x$.
We obtain the scattering rate $\gamma_s$ of pump photons from the Rayleigh scattering cross-section, $\gamma_s = c \sigma_s / V$.

Both rate constants $\gamma_a$ and $\gamma_s$ depend on the dielectric properties of the particle. In case of atomic clusters that are sufficiently large to be modeled as dielectric spheres -- but still small compared to the light wavelength -- the rates can be extracted from the complex polarizability. It reads as \cite{Kreibig1995a,Smirnov2000}
\begin{equation}
 \chi = 4\pi \varepsilon_0 R^3 \frac{\varepsilon (\lambda) - 1}{\varepsilon (\lambda) + 2},
\end{equation}
with $R$ the cluster radius and $\varepsilon(\lambda)$ its complex dielectric function at the wavelength $\lambda$ of the applied light field. This results in the absorption cross section
\begin{equation}
 \sigma_a = \frac{2\pi}{\varepsilon_0 \lambda} \im \chi = 4 \pi \frac{2\pi R}{\lambda} R^2 \im \left( \frac{\varepsilon (\lambda) - 1}{\varepsilon (\lambda) + 2} \right),
\end{equation}
and in the Rayleigh scattering cross section
\begin{equation}
 \sigma_s = \frac{8 \pi}{3} \left( \frac{2\pi R}{\lambda} \right)^4 R^2 \left| \frac{\varepsilon (\lambda) - 1}{\varepsilon (\lambda) + 2} \right|^2 .
\end{equation}
Knowing the dielectric function $\varepsilon(\lambda)$ at the laser wavelength one can compute the scattering and absorption rates, as has been done for the exemplary cluster particles in Table \ref{tab:mol}. Although the resulting rate values there are by orders of magnitude smaller than the cavity decay rate, the momentum diffusion that is associated to the absorption and the Rayleigh scattering of pump photons will generally surpass the diffusion related to the cavity modes, as we will discuss below. In the particular case of sub-wavelength metal clusters the absorption rate is much larger than the elastic light scattering rate, $\gamma_a \gg \gamma_s$. Consequently, the absorption effect will dominate the momentum diffusion along the direction of the pump laser, and the scattering effect may be neglected. This is however not the case in a two-dimensional setting like the one in Fig.~\ref{fig:cav}, where the pump mode and the empty cavity modes are orthogonal to each other. Here, the absorption effect merely causes momentum diffusion along the pump direction $y$, while the diffusion along the $x$-direction is mainly related to the Rayleigh scattering of photons out of the pump beam. Such an experiment would therefore benefit from the much smaller scattering cross sections.

\subsection{Realistic dissipation rates} \label{sec:adddiff}

We have now properly included light absorption and scattering into our master equation model which, after translating those terms into phase space, give us a realistic description of the friction and diffusion dynamics of large particles in a cavity. The translation mainly leads to an additional momentum diffusion that counteracts the envisaged frictional momentum damping in dependence of the respective absorption and scattering rates.

A Rayleigh scattering event can be divided into the momentum recoil due to the absorption of a photon from the pump mode, and the recoil caused by the reemission of the photon. The former part is identical to the effect of genuine photon absorption, i.e. the scattering rate $\gamma_s$ can be added directly to the absorption rate $\gamma_a$ in the absorption Lindblad term (\ref{eqn:absLind}). In a one-dimensional situation this yields the non-negative momentum diffusion
\begin{eqnarray}
 D_{pp}^{\rm (abs)} (x,p) &=& \hbar^2 (\gamma_a + \gamma_s) \left| \alpha \partial_x f(x) \right|^2 \label{eqn:Dabs}
\end{eqnarray}
along the cavity axis $x$. Only $x$-derivatives of the mode function are involved. Consequently, this diffusion term can be suppressed for instance by using a large waist of the pump mode perpendicular to the axis $x$ of a cavity with a number of empty modes.
The latter will cause the desired friction in $x$ according to (\ref{eqn:beta2}), while the pump mode fulfills $\partial_x f \approx 0$ and thus minimizes the additional diffusion term (\ref{eqn:Dabs}). In case of a running wave pump mode $f(x) \notin \R$ directed along the $x$-axis the absorption also leads to a net force
\begin{eqnarray}
 g_{p}^{\rm (abs)} (x,p) &=& 2\hbar (\gamma_a + \gamma_s) |\alpha|^2 \im \left[ f(x) \partial_x f^{*} (x) \right]\label{eqn:gabs}
\end{eqnarray}
that acts on the particle in the preferred direction of the photon flux. It vanishes for a standing-wave pump where the same amount of photons hits the particle from one side and the other.

The reemission part of the scattering results in the diffusion coefficient
\begin{eqnarray}
 D_{pp}^{\rm (sca)} (x,p) &=& \hbar^2 \gamma_s |\alpha|^2 \frac{\omega_p}{c} \left( \frac{\omega_p}{c} \la u_x^2 \ra |f(x)|^2 + 2 \la u_x \ra \im \left[ f(x) \partial_x f^{*} (x) \right] \right) . \label{eqn:Dscatt}
\end{eqnarray}
The angular scattering distribution $N(\vu)$ enters through the average $x$-component of the scattered photon momentum $\la u_x \ra = \int \diff^2 \vu \, N(\vu) u_x$, and the averaged square $\la u_x^2 \ra$. The second term in the diffusion coefficient (\ref{eqn:Dscatt}) is caused by a combined anisotropy of the photon flux of the mode, $f(x) \notin \R$, and of the scattering, $\la u_x \ra \neq 0$. The latter also implies a net force
\begin{eqnarray}
 g_{p}^{\rm (sca)} (x,p) &=& 2\hbar \gamma_s \frac{\omega_P}{c} \la u_x \ra |\alpha f(x) |^2 \label{eqn:gscatt}
\end{eqnarray}
acting on the particle. Both anisotropic forces (\ref{eqn:gabs}) and (\ref{eqn:gscatt}) are independent of the particle velocity and therefore do not contribute to friction.

If the particle scatters photons isotropically into free space, the average $\la u_x \ra = 0$ and $\la u_x^2 \ra = 1/3$.
It is possible to adjust $\la u_x^2 \ra$, i.e. the diffusion due to Rayleigh scattering, by using linear polarized pump light from a mode perpendicular to the cavity axis $x$, for instance. When the polarization axis is parallel to the cavity axis, one obtains $\la u_x \ra = 0$ and $\la u_x^2 \ra = 1/4$ from the corresponding dipolar scattering pattern.

The diffusion terms caused by the absorption and scattering of pump light usually govern the total momentum diffusion of the particle as they exceed the momentum diffusion effect of the cavity modes discussed in Section \ref{sec:phasespace}. In order to compare the respective diffusion terms with the diffusion of the cavity, as given by the low-velocity term (\ref{eqn:Dpp_av1}) we need to average the expressions (\ref{eqn:Dabs}) and (\ref{eqn:Dscatt}) over position. We assume isotropic scattering for the latter term. The average diffusion due to absorption then scales as $(U_0/\kappa)^{-2}(\gamma_a + \gamma_s)/\kappa$ times the cavity diffusion coefficient, while the scattering-induced diffusion scales as $(U_0/\kappa)^{-2}\gamma_s/\kappa$ times the cavity diffusion coefficient. The final kinetic energy that can be reached through the frictional damping process roughly increases by the same factors. Regarding the example of heavy clusters listed in Table~\ref{tab:mol} the increase can be dramatic. For the lithium cluster and the gold cluster the diffusion caused by absorption surpasses the cavity diffusion by a factor of the order of $10^4$. However, as already mentioned, the absorption term can be suppressed by using a pumped mode that is oriented perpendicularly to the axis of interest $x$. In this case, only the diffusion caused by scattering has to be taken into account, which is of the same order of magnitude as the cavity diffusion for all the heavy clusters listed in Tab.~\ref{tab:mol}. Due to the high transparency of silica clusters and helium droplets in the IR regime, they are not influenced by the absorption-induced diffusion even at high masses.

\section{Conclusions}

In this work we have presented a full quantum treatment in terms of master equations of the cavity-assisted motional damping of a polarizable particle. In the limit of weak coupling to a strong laser field we could derive closed expressions for both the velocity-dependent force and the momentum diffusion related to the delayed reaction of high-finesse cavity modes to scattered pump photons. On the single particle level, our phenomenological treatment of the particle-light interaction gives insight into the challenges and peculiarities of optical cooling experiments with ensembles of mesoscopic objects such as large molecules or atomic clusters. Though the weak polarizability of those heavy particles can be compensated by a sufficiently strong pump field intensity, the friction rate has an upper bound of the same order of magnitude as the recoil frequency.
This can be fairly slow but our model predicts that it can also be enhanced by orders of magnitude when using multimode resonator geometries such as a confocal cavity setup. We therefore believe that the presented scheme could prove to be useful in practice, in view of its generality and the lack of alternative cooling schemes for freely moving massive particles. The large multimode enhancement might even be exploited in future optomechanics experiments.
For higher temperature sources containing fast particles we showed that the capture range and region of optimal cooling can be tuned by a suitable choice of detunings and stronger pump. Ultimately, a time-dependent detuning can be used to optimize the capture range and the final temperature for a given interaction time.
The final temperature is however also sensitive to the absorption and the free-space scattering cross-sections of the particle, which should be as small as possible.

Both the frictional part of the force as well as the negativities in the momentum diffusion coefficient are a direct consequence of the memory effect of the cavity, i.e. its delayed reaction towards the field fluctuations caused by the presence of the particle. The memory effect, and in particular the consequential negative diffusion, would play a more major role in the dynamics of the particle beyond the weak coupling limit assumed here. A strictly semiclassical description in terms of an effective Langevin equation of motion for the particle would then have to be replaced by coupled Heisenberg equations for the particle and the field. It is, however, still an open problem to obtain an effective description in terms of a master equation (without a harmonic approximation of the optical potential) for the particle in the strongly coupled regime.

\section*{Acknowledgments}
We thank the Austrian science foundation FWF for support within
the projects SFB1505, I119-N16 (CMMC-Euroquam), and the doctoral
program W1210 (CoQuS).

\appendix

\section{General master equation for $N$ particles in $M$ modes} \label{app:generalME}

We generalize the derivation of the weak coupling master equation for a particle in a pumped cavity mode from Section \ref{sec:me} to $N$ particles in $M$ modes. A single laser frequency $\omega_p$ is used for all stationarily pumped modes. We allow for different pump amplitudes $\alpha_m$, detunings $\Delta_m = \omega_m - \omega_p$, and damping rates $\kappa_m$. In practice one would single out one or at most a few almost degenerate modes out of many to be strongly pumped by a single laser, as described by the master equation (\ref{eqn:ME2}). It is a special case of the equation presented in the following.

The mode functions $f_m (\vr)$ determine the spatial structure of the quadrature field operator
\begin{equation}
 \oE^{+} (\vr)  = \sum_{m=1}^M \sqrt{\frac{\hbar \omega_m}{2 \eps_0 V_m}} f_m (\vr) \oa_m\da,
\end{equation}
which enters the field coupling Hamiltonian (\ref{eqn:H_i}) at each particle's position operator $\ovr_j$. For simplicity, we assume the polarizability $\chi$ to be constant for all modes. In the displaced and rotating frame with respect to the pump the master equation of particles and modes reads
\begin{eqnarray}
 \partial_t \rho &=& -\frac{i}{\hbar} \left[ \oH_C + \oH_P + \oH_I, \rho \right] + \sum_{m=1}^M \kappa_m \left( 2 \oa_m \rho \oa_m\da - \left\{ \oa_m\da\oa_m , \rho \right\} \right), \label{eqn:MEapp}
\end{eqnarray}
with the cavity, particle, and interaction Hamiltonians
\begin{eqnarray}
 \oH_C &=& \sum_{m=1}^M \hbar \Delta_m \oa_m\da \oa_m \\
 \oH_P &=& \sum_{j=1}^N \left( \frac{\ovp_j^2}{2m_P} + \sum_{m,n=1}^M \hbar~U_{mn} \alpha_m^{*} \alpha_n f_m^{*} (\ovr_j) f_n (\ovr_j) \right) \\
 \oH_I &=& \sum_{j=1}^N \hbar \sum_{m,n=1}^M U_{mn} \alpha_m^{*} \oa_n f_m^{*} (\ovr_j) f_n (\ovr_j) + h.c.
\end{eqnarray}
The symmetric coupling matrix $U_{mn} = - \chi \sqrt{\omega_m \omega_n / V_m V_n} / 2\eps_0$ describes the scattering of photons at a particle between mode $m$ and $n$. The dispersive coupling between the strong pump fields is responsible for the optical potential in the particle Hamiltonian $\oH_P$, while the Doppler-shifted scattering of pump photons into a detuned cavity mode at a moving particle may cause dissipation of motion. The latter comes from the interaction term $\oH_I$ after adiabatically eliminating the cavity modes.
The first weak coupling condition in Section \ref{sec:me} stays the same, and has to hold for all modes, $|U_{mn}| \ll \kappa_m$ $\forall m,n$. The second weak coupling criterion however poses a further restriction on the particle number, $|N U_{mn} \alpha_n | \ll \kappa_m$ $\forall m,n$. This is because each particle may scatter photons into the same mode $m$ independently, and the weak coupling limit requires the cumulative photon fluctuations of all particles to be smaller than one.

We may follow the procedure given in Section \ref{sec:me} adapting the Liouvillians there to the corresponding terms in (\ref{eqn:MEapp}). This yields the master equation
\begin{eqnarray}
 \fl \partial_t \rho_P &=& -\frac{i}{\hbar} \left[ \oH_P, \rho_P \right] - \sum_{\ell,m,n=1}^M \sum_{k=1}^N U_{\ell m} U_{mn} \left( \alpha_\ell^{*} \alpha_n \left[ f_{\ell}^{*} f_m (\ovr_k),\og_{mn} \rho_P \right] + h.c. \right),
\end{eqnarray}
with the $N$-particle memory operator
\begin{equation}
 \og_{mn} = \int_{0}^\infty \diff \tau \, e^{-(\kappa_m + i\Delta_m)\tau} \sum_{j=1}^N f_{m}^{*} f_n \left( \ovr_j, \ovp_j; \tau \right).
\end{equation}
The backward-evolved product of the mode functions is given by
\begin{equation}
 f_{m}^{*} f_n \left( \ovr, \ovp; \tau \right) = e^{-i \oH_P \tau / \hbar} f_m^{*} (\ovr) f_n (\ovr) e^{i \oH_P \tau / \hbar}.
\end{equation}

\section{Phase space translation rules} \label{app:wigner}

The Wigner-Weyl phase space formulation is a way to represent observables and density operators of a motional degree of freedom in terms of functions over the phase space variables $(x,p)$. The Weyl symbol of an observable $\oA$ is given by the invertible integral transformation
\begin{equation}
 A(x,p) = \int \diff s \, e^{ips/\hbar} \lla x - \frac{s}{2} \right| \oA \left| x + \frac{s}{2} \rra , \label{eqn:weylsymbol}
\end{equation}
and the Wigner function of a state $\rho$ is its normalized Weyl symbol
\begin{equation}
 w(x,p) = \frac{1}{2\pi \hbar} \int \diff s \, e^{ips/\hbar} \lla x - \frac{s}{2} \right| \rho \left| x + \frac{s}{2} \rra . \label{eqn:wigner}
\end{equation}
It can be seen as the quantum generalization of the classical phase space density. Given the Hamiltonian $\oH = \op^2 /2m_P + V(\ox)$ the Wigner function evolves according to the quantum-Liouville-Equation \cite{Schleich2001}
\begin{equation}
 \fl \left[\partial_t + \partial_x \frac{p}{m_P} - \partial_p V'(x) \right] w(x,p) = \sum_{\ell=1}^\infty \frac{(-)^\ell (\hbar/2)^{2\ell}}{(2\ell + 1)!} \partial_x^{2\ell+1} V(x) \partial_p^{2\ell +1} w(x,p). \label{eqn:QLE}
\end{equation}
The nonclassical behaviour entirely comes from the right hand side, and it represents diffraction effects at the potential. In case of states $w(x,p)$ where the quantum coherence is negligible, the semiclassical approximation is applied by dropping all higher-than-second-order derivatives of the Wigner function. One obtains a Fokker-Planck-type equation (\ref{eqn:FPE})
with the drift terms $g_x$, $g_p$ and the diffusion terms $D_{xx}$, $D_{pp}$, and $D_{xp}$. In a coherent time evolution governed by the conservative force term $g_p(x,p)=-\partial_x V(x)$ there is no diffusion according to the quantum-Liouville-Equation. This is not the case in a general master equation anymore, where the translation into phase space may yield a momentum-dependent dissipative force $g_p (x,p)$, as well as nonzero diffusion terms.

The Weyl symbols of the products $\oA\oB$, $\oB\oA$ of two operators $\oA$, $\oB$ read \cite{Schleich2001}
\begin{eqnarray}
 (AB)(x,p) &=& A \left(x - \frac{\hbar}{2i} \partial_p, p+ \frac{\hbar}{2i} \partial_x \right) B(x,p), \\
 (BA) (x,p) &=& \left[A^{*} \left(x - \frac{\hbar}{2i} \partial_p, p + \frac{\hbar}{2i} \partial_x \right) B^{*}(x,p)\right]^{*},
\end{eqnarray}
where the Weyl symbol with differential operator arguments acting to the right is defined via the Taylor expansion
\begin{equation}
 \fl A \left(x - \frac{\hbar}{2i} \partial_p, p+ \frac{\hbar}{2i} \partial_x \right) = \sum_{n,m=0}^\infty \frac{1}{m! n!} A^{(m,n)} (x,p) \left( - \frac{\hbar}{2i} \partial_p \right)^m \left( \frac{\hbar}{2i} \partial_x \right)^n .
\end{equation}
The Weyl symbols of operator functions $f(\ox)$ and $g(\op)$ are simply given by $f(x)$ and $g(p)$, respectively.

\section{Lindblad terms of photon absorption and scattering} \label{app:abs}

In order to derive a Lindblad term that describes the momentum diffusion due to absorption or elastic Rayleigh scattering of pump photons we utilize and modify the master equation (\ref{eqn:ME2}) of a particle in one pumped and $M$ empty cavity modes.
We assume here the same weak coupling assumptions already made in Sec.~\ref{sec:me}.

In the case of photon absorption we substitute the empty cavity modes by $M \gg 1$ internal undamped oscillators, i.e. we set $f_m (\vr) = 1$ and $\kappa_m = 0$ for all those modes. This follows the textbook analogy of damping, where an oscillator mode is weakly coupled to the infinite free-space heat bath \cite{Scully1999,Gardiner2004,Puri2001}. Consequently, we neglect any heating effect of the internal bath which would lead to additional diffusion by thermal radiation. This is a valid approximation for particles with sufficiently large heat capacities, as considered in this paper. We model the internal bath as a continuum of oscillators with arbitrary detunings from the pump and replace the mode summation in (\ref{eqn:ME2}) by the integral
\begin{equation}
 \sum_{m=1}^M \to \int_{-\infty}^\infty \diff \Delta \, d(\omega_p + \Delta). \label{eqn:continuumapp}
\end{equation}
Here, $d(\omega)$ denotes the internal density of states. The rotating wave approximation around the optical pump frequency $\omega_p$ admits to set the integral bounds to $\pm \infty$. In addition the coupling constant times the density of states is assumed to be flat around the pump frequency,
\begin{equation}
U(\omega_p+\Delta)^2 d(\omega_p+\Delta) \approx  U(\omega_p)^2 d(\omega_p) =: \pi \gamma_a .
\end{equation}
The constant $2\gamma_a$ is the phenomenological rate of absorption events. Plugging everything into Equation (\ref{eqn:ME2}) yields the Lindblad term
\begin{equation}
 \left[ \partial_t \rho_P \right]_{\rm abs} = \gamma_a |\alpha|^2 \left( 2 f (\ovr) \rho_P f^{*} (\ovr) - \left\{ \rho_P , \left| f (\ovr) \right|^2 \right\}  \right).
\end{equation}
It represents the momentum diffusion caused by the absorption of photons from a single pump mode $f(\ovr)$.

A similar calculation can be performed for Rayleigh scattering of light. Starting again from (\ref{eqn:ME2}), we replace the empty modes by the undamped free-space plane waves $f_{\vk} (\vr) = e^{i\vk \cdot \vr}$ with wave vector $\vk = \omega \vu / c$. We disregard the Doppler shift of the elastically scattered pump photons, since we focus on the diffusion effect. The memory operator (\ref{eqn:memop2}) then becomes
\begin{eqnarray}
 \og_{\vk} &=& \int_0^\infty \diff \tau \, e^{-i\Delta \tau} f(\ovr) e^{-i\vk \cdot \ovr} \approx \pi \delta (\Delta) f(\ovr) e^{-i\omega_p \vu \cdot \ovr / c}.
\end{eqnarray}
The principal value due to the semi-infinite integral is dropped as it would only lead to a negligible Lamb shift term \cite{Puri2001,Walls2006}. Given a phenomenological scattering rate $\gamma_s$ and the distribution of scattering angles $N(\vu)$, one arrives at the Lindblad term (\ref{eqn:scattLind}). In the case of Rayleigh scattering of linear polarized pump light the distribution is given by a dipole scattering pattern $N(\vu) \propto \sin \theta$ with $\theta$ the polar angle with respect to the polarization axis.

\section{General friction and diffusion terms} \label{app:generalFPE}

We translate the dissipative part of the general master equation for $N$ particles in $M$ modes from \ref{app:generalME} into phase space, using the translation rules derived in \ref{app:wigner}. We restrict the calculation to the one-dimensional case along the cavity axis $x$. Note that the phase space of $N$ particles is $2N$-dimensional then, with coordinates $(x_1,p_1, \ldots , x_N, p_N)$. A tedious but straightforward calculation leads to the following force and diffusion terms in the semiclassical limit:

The force consists of a first order and a second order contribution in $\hbar$,
\begin{equation}
 g_{p_k}^{\rm (dis)} = g_{p_k}^{(1)} + g_{p_k}^{(2)}
\end{equation}
Both contributions are expressed in terms of the \textit{memory integral per particle}
\begin{equation}
 G_{mn}(x,p) = \int_0^\infty \diff \tau \, e^{-(\kappa_m + i\Delta_m) \tau} f_m^{*}\left( x - \frac{p \tau}{m_P} \right) f_n \left( x - \frac{p \tau}{m_P} \right) . \label{eqn:memintapp}
\end{equation}
The past position in the argument of the mode functions is approximated by the free trajectory $x+p\tau/m_P$, as it is done in Section \ref{sec:phasespace}.

The dissipative force terms now read
\begin{eqnarray}
 \fl g_{p_k}^{(1)} (x_1,p_1, \ldots , x_N, p_N) &=& 2\hbar \sum_{\ell,m,n=1}^M U_{\ell m} U_{mn} \re \left[ i \alpha_\ell^{*} \alpha_n \left( \partial_{x_k} f_\ell^{*} (x_k) f_m (x_k) \right) \right. \nonumber \\
&& \quad \times  \left. \sum_{j=1}^N G_{mn} (x_j,p_j) \right], \\
 \fl g_{p_k}^{(2)} (x_1,p_1, \ldots , x_N, p_N) &=& -\hbar^2 \sum_{\ell,m,n=1}^M U_{\ell m} U_{mn} \re \left[ \alpha_\ell^{*} \alpha_n \left( \partial_{x_k}^2 f_\ell^{*}(x_k) f_m (x_k) \right) \right. \nonumber \\
&&\quad \times \left. \partial_{p_k} G_{mn} (x_k,p_k) \right].
\end{eqnarray}
In the first order term all particles add up homologously via their memory integral to the force acting on particle $k$. The second order force is however only related to the field fluctuations caused by the particle it acts upon.
Finally, the nonzero parts of the diffusion matrices read
\begin{eqnarray}
 \fl D_{p_k p_j} (x_1,p_1, \ldots , x_N, p_N) &=& 2\hbar^2 \sum_{\ell,m,n=1}^M U_{\ell m} U_{mn} \re \left[ \alpha_\ell^{*} \alpha_n \left( \partial_{x_k} f_\ell^{*}(x_k)f_m (x_k) \right) \right. \nonumber \\
&&\quad \times \left. \partial_{x_j} G_{mn} (x_j,p_j) \right], \\
 \fl D_{p_k x_j} (x_1,p_1, \ldots , x_N, p_N) &=& -\hbar^2 \sum_{\ell,m,n=1}^M U_{\ell m} U_{mn} \re \left[ \alpha_\ell^{*} \alpha_n \left( \partial_{x_k} f_\ell^{*}(x_k)f_m (x_k) \right) \right. \nonumber \\
&&\quad \times \left. \partial_{p_j} G_{mn} (x_j,p_j) \right].
\end{eqnarray}
In both diffusion and first order force the particles couple to the field fluctuations caused by the other particles in the same way as they couple to their own field fluctuations. This indicates a strongly correlated system, where one naturally expects collective effects beyond our weak coupling limit, such as self-organizing phase transitions \cite{Domokos2002a,Black2003a}.


\end{document}